\documentclass[apjl]{emulateapj}
\usepackage{hyperref}
\usepackage{color}
\hypersetup{colorlinks,
 citecolor=blue,
 linkcolor=blue}
\usepackage{natbib}

\usepackage{ifthen} 
\usepackage[table]{xcolor}
\usepackage{soul}

\usepackage{hyperref}
\hypersetup{colorlinks,%
 citecolor=blue,%
 linkcolor=blue}
\definecolor{lightblue}{rgb}{.70,.95,1} 
\def\aj{AJ}
\def\apj{ApJ}
\def\apjl{ApJ}
\def\apjs{ApJS}
\def\aap{A\&A}
\def\aaps{A\&AS}
\def\mnras{MNRAS}
\def\nat{Nature}
\def\pasp{PASP}%
\def\araa{ARA\&A}
\def\pasa{PASA}
\def\pasj{PASJ}
\newcommand{\hi}{{\sc H~i}\/ }
\newcommand{\hii}{{\sc H~ii}\/ }

\newcommand{\nii}{{\sc N~ii}\/}
\newcommand{\oiii}{{\sc O~iii}\/}

\def\simlt{\lower.5ex\hbox{$\; \buildrel < \over \sim \;$}}
\def\simgt{\lower.5ex\hbox{$\; \buildrel > \over \sim \;$}}
\def\arcdeg{\hbox{$^\circ$}}
\def\arcmin{\hbox{$^\prime$}}
\def\arcsec{\hbox{$^{\prime\prime}$}}

\begin{document}

\title{MCG+08-22-082: A double core and boxy appearance dwarf lenticular galaxy
  as suspected to be a merger remnant}

\author{Mina Pak\altaffilmark{1,2}, 
Sanjaya Paudel\altaffilmark{1}, 
Youngdae Lee\altaffilmark{1,3}, 
Sang Chul Kim\altaffilmark{1,2,$\dagger$}
}

\affil{\altaffilmark{1}Korea Astronomy and Space Science Institute (KASI), 
   776 Daedukdae-ro, Yuseong-gu, Daejeon  34055, Republic of Korea}
\affil{\altaffilmark{2}Korea University of Science and Technology (UST), 
   217 Gajeong-ro Yuseong-gu, Daejeon 34113, Republic of Korea}
\affil{\altaffilmark{3}Department of Astronomy and Space Science, Chungnam National University, 
   99 Daehak-ro, Daejeon  34134, Republic of Korea}
\email{minapak@kasi.re.kr, sjy@kasi.re.kr, ylee@kasi.re.kr, sckim@kasi.re.kr}

\altaffiltext{$\dagger$}{Corresponding author}

\begin{abstract}
We present a study on the dwarf lenticular galaxy MCG+08-22-082 (U141)
  located in the Ursa Major cluster,
  which is blue centered, double cored, and has a boxy appearance. 
Using publicly available data from the SDSS,
  we perform an analysis of structural and stellar population properties of the galaxy and the cores. 
We find that the light profile of U141 follows an exponential law. 
U141 has a brightness $M_{r} = -16.01$ mag, and an effective radius $R_{e} = 1.7$ kpc. 
The boxiness parameter, $a_4/a$, is mostly between 0 and $-0.05$ in the inner parts,
  reaching an extreme of about $-0.1$. 
Double cores are seen at the center of U141 and each of these cores has
  a stellar mass of $\sim$10$^{6}$ M$_{\sun}$ and
  the separation between them is $\sim$300 pc. 
Optical spectroscopy of these cores shows prominent emission in H$\alpha$ suggesting ongoing star forming activities. 
We interpret these morphological properties to speculate that U141 is a merger remnant of two disk galaxies. 
Thus, we might have caught an intermediate stage of merging with the evidence of double cores in the center of the galaxy. 

\end{abstract}


 \section{Introduction}

Early-type dwarf galaxies (dEs) are the most common galaxy type
  in clusters of galaxies \citep{Jer97}.
Recently, many studies of dEs have revealed that
  they possess diverse substructural features 
  such as spiral disks, bars, nuclei and/or central star forming regions 
  \citep{Jer00, DeR03, Lis06a, Lis06b, Janz14}. 

Galaxy morphologies are one of the fundamental elements 
  to understand their underlying dynamics and assembly histories. 
Several formation scenarios are proposed that integrate the variety of observable features of dEs. 
The most discussed hypothesis involves transformation from late-type dwarf galaxies
  into dEs with the surrounding environments playing a significant role. 
Moreover, owing to their shallow potential wells \citep{deB08},
  dwarf galaxies are fragile to the environmental mechanisms.
Ram pressure stripping \citep{Gun72} and harassment \citep{Moo98} are the most favored mechanisms
  generating morphological transformation from late-type galaxies into dEs in the harsh 
  environment. 
\citet{Pen14} presented a study on an interesting barlike dE, SA 0426-002,
  in the core of the Perseus cluster. 
They argued that SA 0426-002 is being tidally transformed into an early-type morphology
  via galaxy harassment, as it is located near NGC 1275, the brightest cluster galaxy. 

In addition to the environmental effects, morphological transformation by merger is
  a generally accepted concept for the formation of large elliptical galaxies
  as demonstrated by simulations \citep{Bar92}. 
In fact, the well known cosmology, i.e $\Lambda$CDM, predicts
  a hierarchical growth of large scale structures, and it is commonly expected that
  galaxies may have undergone through several major and/or minor merging episodes.

\cite{Nab06} performed a numerical simulation of an idealized equal mass merger of galaxies
  and showed that the final remnant galaxies are slow rotating and boxy,
  with double peaks in their center as well (see figure 11 of \citealt{Nab06}). 
Generalizing these results indicates that boxy morphology and double cores are the signs of mergers,
 which are commonly seen in large elliptical galaxies (\citealt{Bek97}; \citealt{Tad12}). 

As an observational evidence of the mergers of dwarf galaxies,
  \citet{Gra12} presented an exotic dE, LEDA 074886, with an overall rectangular shape. 
They suggested that this galaxy is fast rotating ($v_{rot}/\sigma$ $\sim$ 1.4)
  which is in accordance with the results of merging of two disk galaxies \citep{Hof10}. 
A comprehensive analysis by \citet{Janz14} of structural parameters of a large sample of dEs
  in the Virgo cluster reveals that disks in low mass galaxies might be associated with
  both disky and boxy appearances. 
They suggest that boxy shapes might be able to be formed by encounters and mergers
  of dwarf galaxies.
A dE with a rectangular shape and double cores has not been reported yet,
  and in this paper we report one of such galaxies.

Our focus is on MCG+08-22-082, a galaxy having a unique morphology with a boxy isophote,
  a blue center, and double cores (Figures 1 and 2). 
This galaxy is a member of the Ursa Major cluster (\citealt{Tre02}; \citealt{Kar13}; U141, \citealt{Pak14}). 
Hereafter, we refer to MCG+08-22-082 as U141 for convenience. 
U141 is classified as a dwarf lenticular galaxy with an absolute magnitude of
  $M_{r} = -16.01$\footnote{Distance modulus of the Ursa Major cluster is 
  $(m-M)_0 = 31.20$ (d = 17.4 Mpc) \citep{Tul12}.} mag. 
The Ursa Major cluster comprises diffuse environments with low velocity dispersion, and thus the tidal interactions and merging between members can occur more frequently than in dense clusters \citep{Mam90}. 
We will investigate a plausible mechanism forming this galaxy using multi-band photometric and spectroscopic data. 

We present a detailed structural and stellar population analysis of U141 
  using the publicly available archival data from the Sloan Digital Sky Survey (SDSS)
  and the Galaxy Evolution Explorer ($GALEX$). 
This paper is organized as follows. 
Section 2 describes the data adopted from the SDSS and
  how we performed surface photometry and spectral analysis. 
Our conclusion and discussion on the evolutionary path of U141 is given in section 3.

\section{The Data and Analysis}  
\subsection{Surface Photometry}  

\begin{table*}     
\caption{Basic information on U141}
\label{phot}
\begin{tabular}{ccc}
\hline
Parameter & Information & Reference \\
\hline
Galaxy name & MCG+08-22-082, U141, & [1, 2] \\
  & PGC 038825, SDSS J121122.56+501610.2 &  [1] \\
R.A. & $182.8448\arcdeg$ ($12^{\rm h}$ $11^{\rm m}$ 22.5$^{\rm s}$, J2000) & [1] \\
Dec. & $+50.2696\arcdeg$ ($+50\arcdeg$ $16\arcmin$ $10\arcsec$, J2000) & [1] \\
$l$, b & 138.557908\arcdeg,   65.625904\arcdeg & [1] \\
$u, g, r, i, z$ & $17.28\pm0.04$, $15.76\pm0.02$, $15.19\pm0.02$, $14.88\pm0.02$,
    $14.93\pm0.03$ mag & [2] \\
$M_{r}$ & $-16.01$ mag & [2] \\
$R_{e}$ & 1.7 kpc & [3] \\
$\mu_{e}$  & 25.54 mag arcsec$^{-2}$ & [3] \\
S\'ersic index, $n$ & 1.3 & [3] \\
Redshift, z & 0.003 & [1] \\
Radial velocity, $v_r$ & 899 km s$^{-1}$ & [1] \\
Total star formation rate, log SFR$_{FUV}$ [M$_{\sun}$ yr$^{-1}$] & $-2.85$ & [4] \\
\hline
\end{tabular}
\\
\\
References are [1] NASA/IPAC Extragalactic Database, 
[2] \cite{Pak14}, 
[3] This study (SDSS $r$-band image), [4] This study (using the GALEX FUV flux).
\end{table*}

\begin{figure}
\includegraphics[width=9cm]{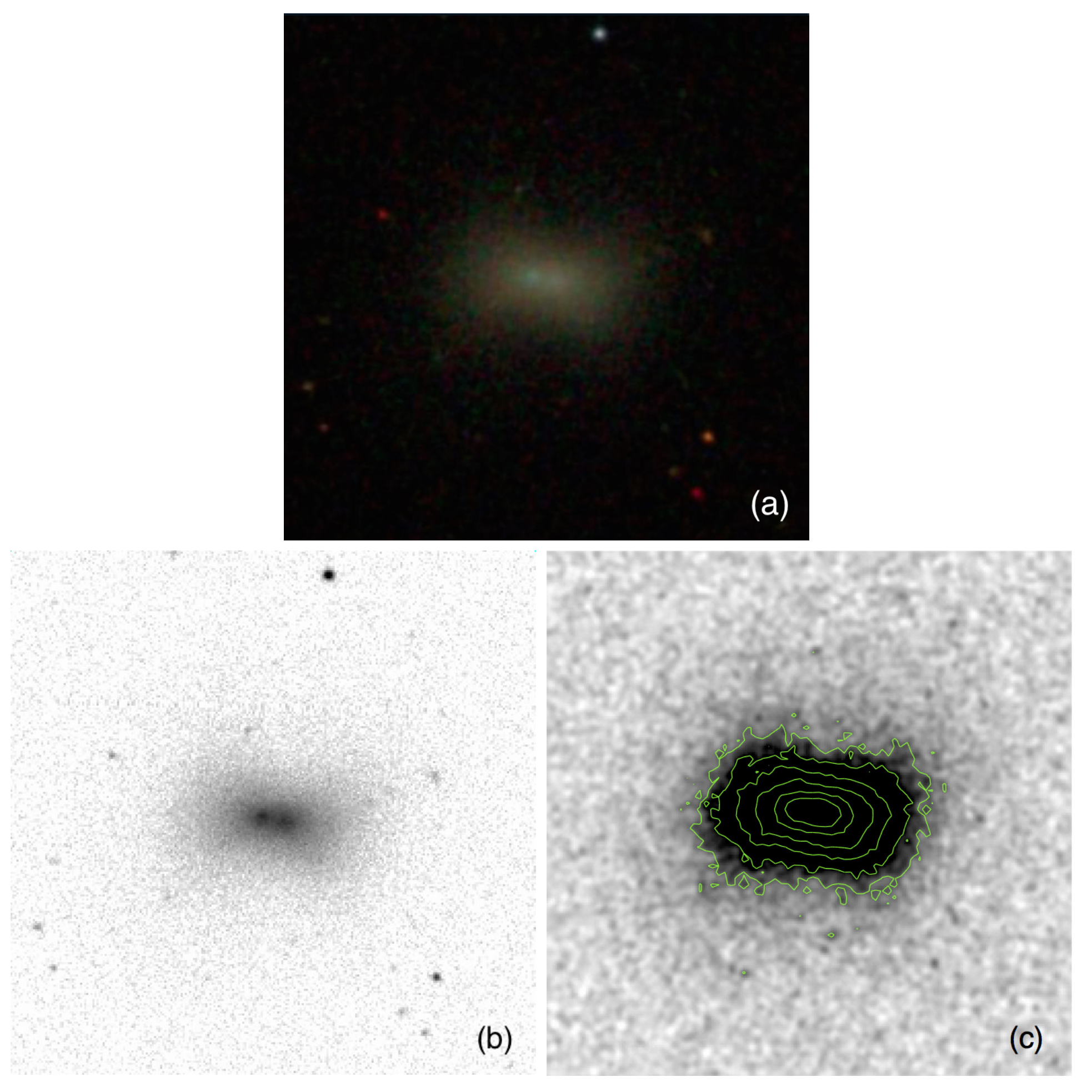}
\caption{(a) A color image cutout from the SDSS skyserver.
(b) SDSS $g$-, $r$-, and $i$-band combined image (see text for details) to increase
  the signal-to-noise ratio compared to separate $g$-, $r$-, and $i$-band images.
(c) Contours are overlapped on thumbnail image (b) from ds9 image tool. 
For all images, size is $100\arcsec \times 100\arcsec$ each and north is at the top and east is to the left.
\label{F1}}
\end{figure}

\begin{figure}
\includegraphics[width=8.5cm]{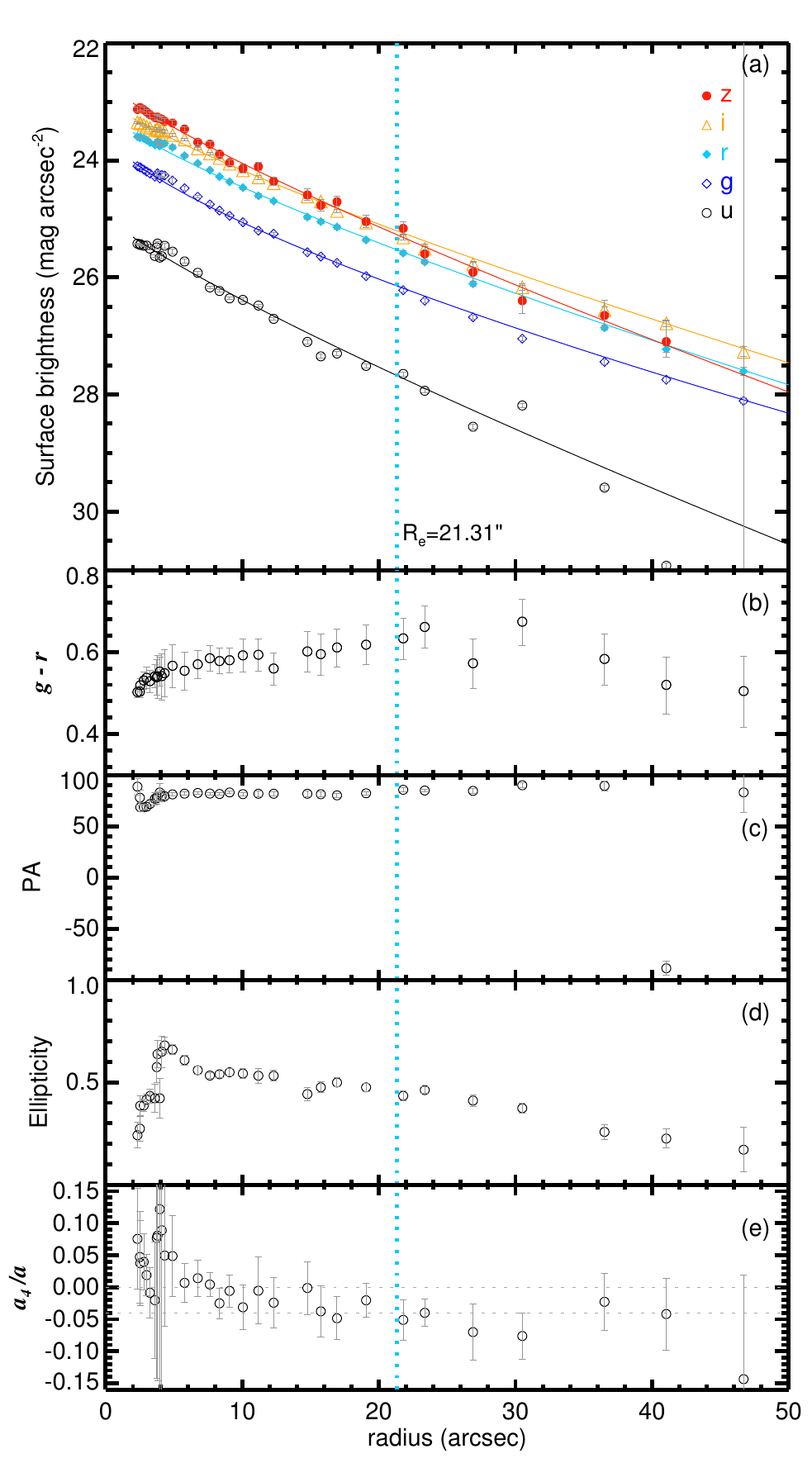}
\caption{Surface brightness and color profiles (panels a and b) and 
  variations of isophotal parameters (panels c to e) as a function of radius are shown. 
The error bars are significantly large at the outer parts with radius larger than $50\arcsec$. 
Blue vertical dotted lines in all panels imply the effective radius in $r$-band obtained by using
  the IRAF/{\tt ELLIPSE} task.
(a) Surface brightness profiles for SDSS $u$-, $g$-, $r$-, $i$-, and $z$-band images
  as a function of radius. 
The solid lines are the best fit S\'ersic function in each filter. 
(b) $g-r$ color profile. (c) Position angle profile. (d) Ellipticity profile. 
(e) The $a_4/a$ profile, where two dotted lines are shown. 
The upper line is limit with the value of zero and the lower line is for the value of $-0.04$,
  which is the most extreme one boxy galaxies can have as suggested by \citet{Hao06}. 
\label{F2}}
\end{figure}

To perform detailed surface photometry on U141, we extensively use imaging from the SDSS Data Release 7
  (DR7, \citealt{Aba09}). 
Reduced and calibrated CCD images were acquired from the SDSS data archive server. 
These were observed in five optical bands, i.e. $u$, $g$, $r$, $i$, and $z$, with a total exposure time of 54s in each band. 
The pixel scale is $0.396\arcsec$ pixel$^{-1}$ and the average seeing in this field of view is $1.2\arcsec$ in the $r$-band
  as we have measured from the Full Width at Half Maximum (FWHM) of randomly selected foreground stars. 
Details of post-processing of the SDSS images are given in \cite{Pak14}. Since sky background subtraction in the SDSS pipeline processed images is not reliable \citep{Lis06a}, we optimized the post-processing steps to improve reliability. In this work, we use the sky background subtracted postage images with an area of 800$\times$800 pixels.

Figure \ref{F1} shows the color image cutout from the SDSS skyserver (panel (a)) and
  the SDSS $g$-, $r$-, and $i$-band combined image of U141 in panel (b)
  with a size of $100\arcsec \times 100\arcsec$. 
Following the scheme and weight factors (equation (1) in \citet{Kni04}) for each band,
  we combined the SDSS $g$-, $r$-, and $i$-band images to increase the signal-to-noise ratio
  by a factor of $\sim$$\sqrt 3$ compared with that in the $r$-band image alone \citep{Lis06a}. 
Double cores and a boxy shape are clearly seen in Figure \ref{F1} (b) and (c), respectively. 
In Figure \ref{F1} (b), we show a contrast stretched version to make the double cores more visible at the center of the galaxy. 
In Figure 1 (c), where we overlaid the contours, three-pixel smoothing has been applied
  to enhance the signal at the outer part of the galaxy.

Basic information on U141 is summarized in Table \ref{phot}. 
Brightness and the coordinates are adopted from Table 2 in \citet{Pak14} and
  from the NASA/IPAC Extragalactic Database. Structural parameters are obtained from surface photometry (see below). 
We calculate the total star formation rate (SFR) from the FUV flux
  measured in \citet{Pak14} and using equation (3) from \citet{Lee09}.
  
We derive a major axis light profile fitting the elliptical isophotes in the images of all bands. 
The IRAF/{\tt ELLIPSE} \citep{Jed87} task is used for this purpose,
  in which the sky-background subtracted images were used and
  unrelated foreground and background objects
  were masked manually. During the ellipse fitting, we fix the center,
  while the position angle and the ellipticity of isophotes are allowed to vary.
We provide initial values of these isophotal parameters as returned from
  the Source Extractor (SE{\sc xtractor}; \citealt{Ber96}) photometry. 
We separately measure the surface brightness profiles
  in the images of all bands using the same initial parameters.

\begin{table}   
\caption{Results of fits to the surface brightness profiles}
 \label{tbl2}
\begin{tabular}{ccccccc}
\hline
Band  & $\mu_{0}$  & $\mu_{e}$ & $R_{e}$ & $n$ \\
          & mag arcsec$^{-2}$ & mag arcsec$^{-2}$ & arcsec &\\
\hline
$u$ & 24.94$\pm$0.12 & 27.17$\pm$0.12 & 16.59$\pm$1.64 & 1.18$\pm$0.12 \\ 
$g$ & 23.54$\pm$0.02 & 26.23$\pm$0.02 & 22.34$\pm$0.25 & 1.40$\pm$0.02 \\
$r$ & 23.17$\pm$0.02 & 25.54$\pm$0.02 & 21.31$\pm$0.22 & 1.25$\pm$0.02 \\
$i$ & 22.96$\pm$0.02 & 24.87$\pm$0.02 & 22.02$\pm$0.27 & 1.23$\pm$0.02 \\ 
$z$ & 22.65$\pm$0.06 & 25.25$\pm$0.06 & 17.90$\pm$0.92 & 1.20$\pm$0.07 \\
\hline
\end{tabular}
\end{table}

Figure \ref{F2} shows the results. 
We present the surface brightness profiles of U141 in all the five filters in panel (a),
  in which we overplot $u$-, $g$-, $r$-, $i$-, and $z$-band surface brightness profiles
  from bottom to top. 
As expected, the $r$-band data points have the smallest error bars,
  and therefore we primarily use the $r$-band surface photometry
  to derive the structural parameters of U141. 

The observed surface brightness profile $\mu(R)$ is fitted with a S\'ersic function
  of the form given in
  \begin{equation} \mu(R)=\mu_0 + \frac{2.5} {\ln(10)} \left(\frac{R}{h}\right)^{1/n}, \end{equation}
  where $\mu_0$ is central surface brightness, $h$ is scale-length,
  and $n$ is the S\'ersic index \citep{Gra05}. 
This process gives the values of $\mu_0$, $h$, and $n$ as output. 
An efficient curve fitting code, $MPFIT$, which is implemented 
  in the IDL library has been used to obtain best fit parameters in each band \citep{Mark09}. 
The inner $5\arcsec$ is excluded during the fit to avoid the central double cores. 

The effective radius $R_{e}$ is obtained using the equation
\begin{equation}
R_e = b^n h,
\end{equation}
  where $b = 1.9992n-0.3271$, for $0.5 < n < 10$,
  and the effective surface brightness, $\mu_{e}$, is calculated by using
\begin{equation}
\mu_0 =\mu_e - 2.5 b / \ln(10)
\end{equation}
  \citep{Gra05}.

All the derived structural parameters in each band are summarized in Table \ref{tbl2}. 
Each column presents (1) band, (2) central surface brightness, (3) effective surface brightness,
  (4) effective radius, and (5) S\'ersic index. 
The errors of surface brightness profiles are measured from the rms scatter of intensity
  along the ellipse fit.
These errors are used as weight values to calculate the errors of the fitted S\'ersic parameters
  using $MPFIT$ package which measures errors from the covariance metrics.
We find that U141 has a nearly exponential light profile with the best fit S\'ersic index 
  equal to 1.3 for the $r$-band. 
This is consistent with the results from other bands. 
The $r$-band effective radius is $R_e = 21.31\arcsec$ (1.7 kpc\footnote{We use a conversion factor
  $1\arcsec = 80$ pc to convert the angular to physical scale.}),
  which is consistent with the results from the $g$- and $i$-bands,
  but is much larger compared to the values from the $u$- and $z$-bands. 
Effective surface brightness in the $r$-band is $\mu_{e,r} =25.54$ mag arcsec$^{-2}$.
 
For the purpose of obtaining the radial $g-r$ color profile, 
  we match the image quality in the $g$- and $r$-bands. 
FWHM estimates of the $g$- and $r$-bands are $1.3\arcsec$ and $1.2\arcsec$, respectively. 
We therefore degrade the $r$-band image by smoothing with a Gaussian kernel of 0.6 pixels. 
We consider the point between the cores as the photometric center. 
Finally, we derive the color distribution from the azimuthally averaged light profiles 
  in the $g$- and $r$-bands. 
In Figure \ref{F2} (b), we show the $g-r$ color profile of U141; 
  here we can clearly see that, as in the color image,
  a positive color gradient is eminent where $\Delta(g-r)$ is 0.18 mag within the R$_e$. 
The color profile becomes nearly flat beyond 10$\arcsec$ with an average $g-r$ color index of 0.6 mag.

We show the variation of the isophotal parameters of position angle (PA), 
  ellipticity ($e$) and boxiness parameter ($a_4/a$) in the third, fourth and
  fifth panels of Figure \ref{F2}, respectively. 
All three parameters remain nearly unchanged beyond 10$\arcsec$ where the average values of PA and
  ellipticity are 79$\arcdeg$ and 0.5, respectively. 
The {\tt ELLIPSE} task measures isophote deviations and gives amplitudes
  using a Fourier series shown in
\begin{equation}
{I(\theta) = I_{0} + \Sigma_{n=1}^{\infty} (A_n \cos n\theta + B_n \sin n\theta),} 
\end{equation}
where $\theta$ is the azimuthal angle and $I_{0}$ is the average intensity over the ellipse. 

The $A_n$ and $B_n$, given by the {\tt ELLIPSE} task,
  are the higher order Fourier coefficients
  divided by the semi-major axis length, $a$, and the local gradient along the major axis 
  \citep{Hao06}.
As a boxiness parameter, $a_4/a$ is defined as $\sqrt{(1-e)} A_4$
  (see \citet{Hao06, Mil99} for details)
  where $A_4$ from the {\tt ELLIPSE} task is a measure of the isophotal deviation
  from an ellipse\footnote{$A_4$ is the Fourier coefficient used in the studies of bars,
  where it is always positive (see, e.g., \citealt{Lau06}).}.
Negative and positive values of $a_4/a$ mean boxy and disky shape of a galaxy,
  respectively (\citealt{Ben88}, \citealt{Hao06}). 
In Figure 2 (e), $a_4/a$ of U141 shows a variation from $-$0.15 to 0.15. 
Interestingly, while the value of $a_4/a$ is mostly between 0 and $−$0.05 
  within 10$\arcsec$, even reaching about $-$0.1 at the extreme,
  the average value of $a_4/a$ beyond 10$\arcsec$ radius is $-$0.06. 
This implies that the outer isophotes of U141 might show boxy shapes.

\subsection{Properties of cores}    

\begin{table*}  
\caption{Properties of the two cores}
\label{cort}
\begin{tabular}{ccccccc}
\hline
 Core & $M_{r}$ & $g-r$ & M$_{*}$& SFR & 12 $+$ log(O/H) & $v_{r}$\\
   & mag & mag & 10$^{6}$ M$_{\sun}$ & 10$^{-3}$ M$_{\sun}$ & dex & km s$^{-1}$ \\
   \hline
   A & $-11.08$ & $-0.22$ & 0.5 & 0.7 & 8.4 & $781 \pm 2.6$ \\
  B & $-10.97$ & 0.10    & 1   & 0.1 & 8.6 & $778 \pm 8.8$ \\
   \hline
 \end{tabular}
 \\
 \end{table*}

\begin{figure}  
\includegraphics[width=8.5cm]{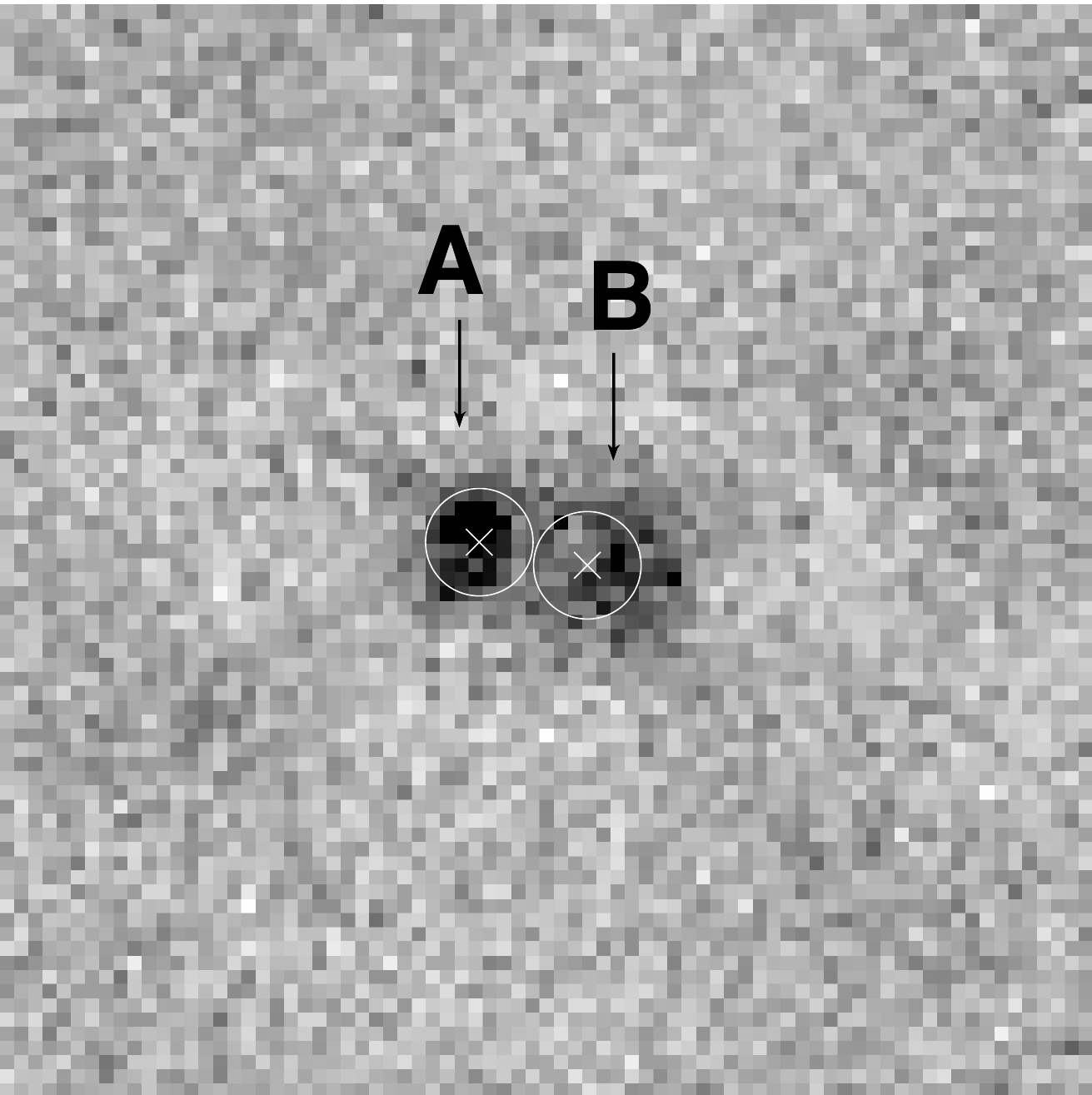}
\caption{Zoom in view of SDSS $r$-band image center of U141, where two cores are distinctly visible. 
The crosses represent the positions of SDSS fibers and the circles are their diameters (each $3\arcsec$).
The size of the field is $30\arcsec \times 30\arcsec$, and north is at the top and east is to the left.
}
\label{core}
\end{figure}

At first glance U141 is a typical blue centered dE,
  as \cite{Lis06b} have identified in the Virgo cluster. 
But, interestingly, a careful inspection of the SDSS images shows that
  U141 possesses double cores and
  they are notably bluer than the main body of the galaxy
  (see Section 3.1 for the discussion on this
  in the context of the merger scenario).

To perform photometry and measure the stellar masses of these cores,
  we subtracted a smooth galaxy model that
  we had obtained from surface photometry in the previous section. 
In the SDSS $r$-band image, the centroids of the two cores are separated by 10 pixels,
  corresponding to a physical separation of 3.9$\arcsec =$ 312 pc. 
While we consider the center of the fitted ellipses in the outer part (i.e. beyond 10$\arcsec$)
  to be the global photometric center of the galaxy,
  this does not coincide with the positions of either of the two cores. 

We show the galaxy subtracted residual image in Figure \ref{core}, in which
  the cores are named as A (east) and B (west). We perform aperture photometry to measure the total flux. 
During the measurement of the flux of each core, we first mask the other and
  use a circular aperture of 10 pixels. 
The sky background is selected from an annulus of 11 pixels and 15 pixels 
  as the inner and outer radii, respectively. 
We find that the two cores are similar in brightness, but
  the core A is slightly bluer and more compact than B; 
  we can also see this in the SDSS color image (Figure \ref{F1} (a)). 
Using the formula provided by \cite{Bell03}, we derive rough stellar masses of these cores 
  from the $r$-band luminosities and $g-r$ colors. 
The results are listed in Table \ref{cort}.
Since the mass to light ratio obtained from \cite{Bell03} heavily depends on color only,
  the redder core B seems more massive than core A.
However, these mass estimates should be taken as rough estimates,
  not exact values.

We find that the SDSS has targeted U141 twice placing the fiber of each core in different epochs. 
The spectrum of B is already available in DR7 and later DR12 adds that of A. 
The SDSS spectra are obtained with fibers of 3$\arcsec$ diameter. We show positions of fibers on U141 in Figure \ref{core}, where the crosses represent the positions and circles the fiber area. 
It seems that the fibers are fairly well centered at the centroid of the cores. 
There is almost no overlap between them but given the typical seeing of $\sim$1$\arcsec$ of SDSS observations, it is likely that a slight fraction of fluxes might have been exchanged.

\begin{figure}  
\includegraphics[width=9cm]{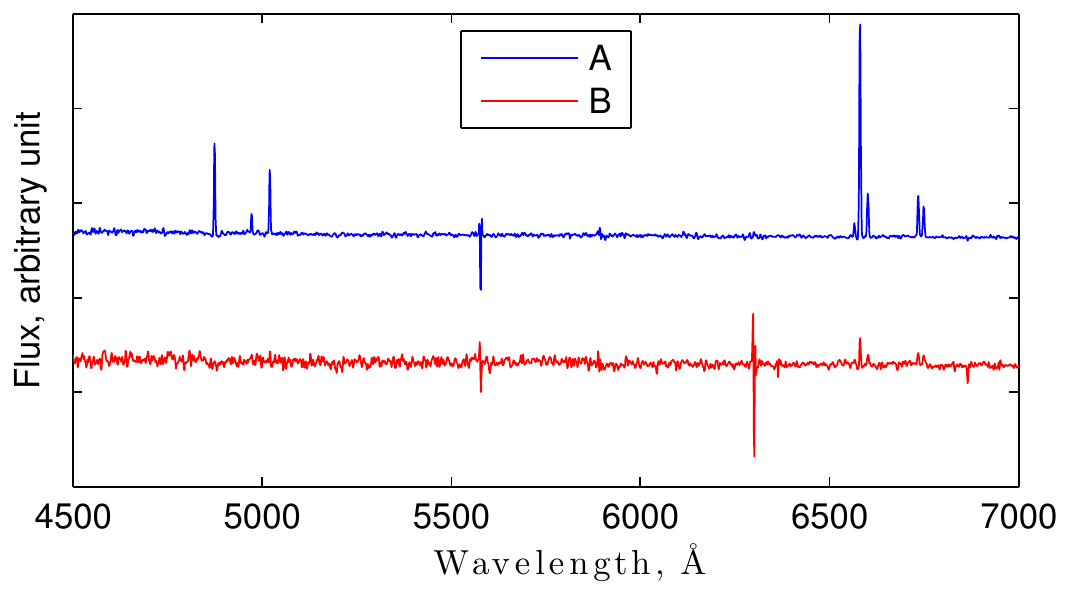}
\caption{SDSS optical spectra of cores A (blue, upper spectrum) and B (red, lower one). }
\label{spec}
\end{figure}

The optical spectra of these cores exhibit the emission lines of typical \hii regions (Figure \ref{spec}). 
The radial velocities of the A and B cores are $781 \pm 2.6$ km s$^{-1}$ and
  $778 \pm 8.8$ km s$^{-1}$, respectively. 
This difference is well within the uncertainty of the radial velocities as measured
  from the SDSS spectroscopy. 
The emission line fluxes are measured after subtracting the stellar absorption features using the publicly available IDL code GANDALF \citep{Sarzi06} and the stellar templates of \cite{Tremonti04}.

Gas-phase oxygen abundances, 12 + log(O/H), for the two cores were estimated
  using two methods described, among others, by \cite{Mar13}; the so-called N2 and O3N2 methods. 
The N2 method only considers the line ratio between H$\alpha$ and [\nii]
  while the O3N2 method uses a combination of the line ratios H$\alpha$/[\nii] and [\oiii]/H$\beta$. 
We obtained 12 + log(O/H) = 8.4(8.3) and 8.6(8.5) dex from the N2(O3N2) method
  for the A and  B cores, respectively. 
The systematic error of these two methods is 0.2 dex. 
We measure the star formation rate from H${\alpha}$ emission flux and calibration provided by \cite{Ken98}. 
For the A and B cores, the star formation rates are $0.7 \times 10^{-3}$ and
  $0.1 \times 10^{-3}$ M$_{\sun}$ yr$^{-1}$, respectively. 
The sum of the star formation rates in the two cores derived from H$\alpha$ emissions
  is nearly half of the global value derived from total FUV emission. 
The H$\alpha$ equivalent widths are 83 \AA\, and 6 \AA\, for the A and B cores, respectively.

\subsection{Environments}  
\begin{figure}   
\includegraphics[width=9cm]{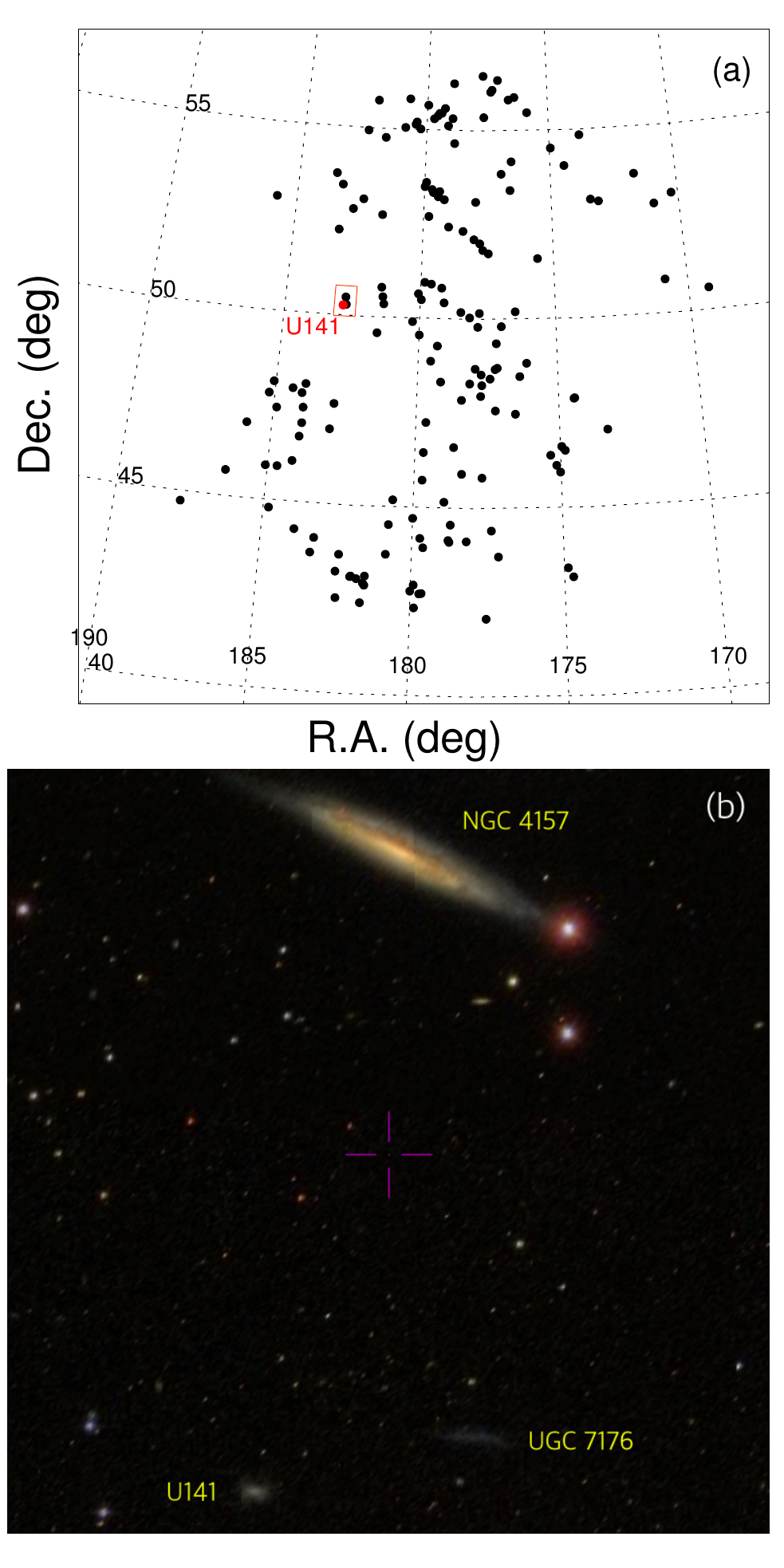}
\caption{Location of U141 in the Ursa Major cluster \citep{Pak14} is marked in panel (a). 
Black dots are all members of the Ursa Major cluster and U141 is shown as a red dot. 
Panel (b) is a magnified image around U141 (red box in panel (a)) from SDSS made by using Aladin imaging tool. 
Magenta cross is the image center with R.A.(J2000) = $12^h$ $11^m$ $05.58^s$ ($182.77325\arcdeg$) and Dec. (J2000) = $+50\arcdeg$ $22\arcmin$ $59.3\arcsec$ ($+50.383139\arcdeg$).
Image size is $18\arcmin \times 18\arcmin$.
U141 lies 23 kpc and 67 kpc apart from UGC 7176 and NGC 4157, respectively. }
\label{env}
\end{figure}

Figure \ref{env} presents the spatial distribution of galaxies in the Ursa Major cluster field and
  neighbors of U141 to show their environments. 
U141 is located in the middle of the cluster (Figure \ref{env} (a)) but not in the dense region. 
Within the sky projected radius of 100 kpc, there are only two close neighbors,
  UGC 7176 (U138) and NGC 4157 (U140). 
U141 has relative radial velocities of 131 km s$^{-1}$ and 207 km s$^{-1}$
  to UGC 7176 and NGC 4157, respectively. 
NGC 4157 is a bright S0 galaxy and UGC 7176 is a dwarf irregular galaxy
  with magnitudes $M_{r}=-20.30$ and $-15.47$ mag, respectively.

We carefully inspect the SDSS $r$-band image of this region to search for
  any tidal features (i.e., stellar stream or filament) that may have originated from
  the past interaction among galaxies. 
Within the detection limit of the SDSS, no such feature is observed
  around either U141 or its neighboring galaxies \citep{Kre12, Kim12, Hir13}.

\section{Conclusion and Discussion}

In this work we have presented a study of structure and stellar populations study in a double core boxy dwarf galaxy, U141. U141 is a member of the Ursa Major cluster, located near the bright S0 galaxy NGC 4157. 
Using imaging and spectroscopy data available from the SDSS, we have performed an analysis of structural and stellar population properties. 
The absolute magnitude and the effective radius of U141 are $M_r = -16.01$ mag and $R_e = 1.7$ kpc, respectively. 
We find that the surface brightness profile is nearly exponential and
  the outer isophotes are clearly boxy, as all the values of $a_4/a$ at radius
  larger than $10\arcsec$ are negative.

Overall, the central region of U141 is bluer than the outer part,
  and in the central area we also identify two distinct compact star-forming cores. 
These two cores are clearly visible in both of the SDSS $r$-band and color images. 
The east core, A, shows strong star forming activity, which is consistent with the prominent emission of H$\alpha$ with equivalent width of 83 \AA. 
Each core has stellar mass of $\sim$10$^{6}$ M$_{\sun}$.

Using the GALEX UV and the SDSS optical data, we carry out a multi-wavelength study of U141.
We perform two-component SED fitting \citep[as described in][]{Jeo07}
  based on the simple stellar population model of \cite{Bru03}. 
We find that U141 is composed of young ($\sim$0.8 Gyr) and old ($\sim$12 Gyr) components that comprise $\sim$3\% and $\sim$97\% of total mass, respectively. 
The 3\% fraction of the young population is significantly larger than that of the total core mass
  that we have derived from different methods. 
The total core mass amounts to only $\sim$0.4\% of the total stellar mass of U141
  ($\sim 3.9 \times 10^8$ M$_{\sun}$). 
To explore further, we visually examine the optical $u$ and UV images and we notice that
  the young stellar populations are not only concentrated in the cores but also
  spread much farther out, although it seems that the cores contain a significantly larger fraction. 
This may explain the observed discrepancy in the young stellar populations in cores
  and the galaxy overall. 
However, it should be noted that the stellar mass calculation
  using very limited photometric data points is highly uncertain and model dependent.

\subsection{A Possible Merger Origin}   

With the discovery of morphological diversity in dEs by \citet{Lis06a,Lis06b},
  our understanding on the formation and evolution of these low mass objects has become even more complicated. 
A considerable fraction of dEs possess substructural features
  (i.e., blue center, disk spiral arm, bar and/or central nucleus), and
  such substructural varieties are inexplicable with a single formation scenario of dEs.
The role of the environment, which can act in several
  different ways \cite[see the reviews][]{Bos06,Lis09}, has been frequently discussed in the literature. 
Although the strong environmental effects could be thought
  as the role for various morphologies of dwarf galaxies,
  another channels for the morphological diversity in dEs could be suggested.

The star clusters or super star clusters in a galaxy could be observed
  as double cores, just like those embedded in U141.
It is possible that the star clusters go into the centers of galaxies due to the dynamical friction
  and the sizes of the cores and star clusters might be similar.
Mean effective radii of star clusters are $3-5$ pc ($0.038\arcsec-0.063\arcsec$)
  at the distance of the Ursa Major cluster \citep{Seth06}. 
This size can not be resolved in our SDSS $r$-band image with seeing of 1.2$\arcsec$. 
In the $r$-band, the FWHMs of cores A and B are 8.1$\arcsec$ and 4.3$\arcsec$, respectively. 
The FWHMs of double cores are large enough to be resolved at the distance of the Ursa Major cluster,
  so these cores might not be star clusters or super star clusters.

Another possibility for the origin of the double cores is that they could be star forming \hii regions. 
\citet{Gu06} presented two distinct nuclei of IC 225 in the off-center region,
  which is somewhat similar to what we observed in U141. 
They suggested that the off-center nucleus is originated by star forming activity
  with high metallicity (12 $+$ log(O/H) $=$ 8.98). 
The metallicity measured in this region is higher than that (12 $+$ log(O/H) $=$ 8.2) 
  calculated by using the luminosity-metallicity relation of isolated dwarf galaxies \citep{Duc04}
  at $M_B = -17.14$.
In the case of U141 ($M_B \sim -15.03$\footnote{From $B = 16.17$ mag, which is calculated
  using the magnitude transformation equation of Lupton (2005) 
  ($B = g + 0.3130\times (g - r) + 0.2271$) in 
  http://classic.sdss.org/dr4/algorithms/sdssUBVRITransform.html .}), the cores of U141 have
  higher metallicities (12 $+$ log(O/H) $=$ 8.4 and 8.6) than that (12 $+$ log(O/H) $\sim$ 8.0)
  of isolated galaxies in the luminosity-metallicity relation \citep{Duc04}.
This implies that the cores might be star forming \hii regions.

\citet{Deb06} found double nuclei with age of $\sim$8 Gyr at the center of
  dwarf elliptical galaxy VCC 128, of which absolute magnitude is $M_B = -15.5$. 
To maintain the observed configuration of double nuclei with a small separation of $\sim$32 pc,
  they introduced a super massive black hole without any direct or indirect evidences. 
If the double nuclei are nuclear disks surrounding a super massive black hole
  with $\sim 10^6$ M$_\odot$, they would be a stable, long-lived configuration and
  would account for the old populations. 
In the case of U141, the separation of the double cores is 312 pc. 
This is much larger than the Bondi radius (14 pc) of a black hole \citep{Bon52}
  where the matters are gravitationally bound to and are being accreted into the black hole
  assuming that the mass of the black hole is $\sim 10^6$ M$_\odot$
  in the middle of the double cores and
  a relative velocity of $\sim$25 km s$^{-1}$ between the black hole and cores
  which is velocity dispersion of a galaxy with M$_B \sim -15.0$ \citep{DeR05}. 
The separation of the double cores in the central massive black hole hypothesis is
  too large to accrete matter into the central black hole. 
Therefore, U141 might not have a massive black hole.

Galaxy merger could be a possible mechanism to form the double cores. 
Double cores are frequently observed in giant early-type galaxies
  with disturbed features (\citealt{She12}; \citealt{Nes15}). 
Dwarf galaxies with double cores are rarely reported. 
It seems that U141 is located in a good condition to have a merge possible, i.e.,
  in a low-density cluster with a low velocity dispersion. 

Boxy isophote dEs are rarely observed. 
LEDA 074886 is an almost rectangular-shaped dE,
  while its surface brightness profile is exponential. 
\cite{Gra12} suggested a hybrid formation scenario that LEDA 074886 might have emerged
  through both dissipational merger in the center and
  dry merger in the outer part of the galaxy. 
\cite{Kaz11} performed simulations of formation of dEs by merging of two dwarf disk galaxies. 
According to their results, boxy morphology of U141 could be formed by both major and minor mergers.
  
The fact that U141 shows a mixture of features with double cores and boxy-shaped morphology
  in the Ursa Major cluster (low cluster velocity dispersion and no intra cluster medium)
  prefers to dwarf-dwarf merger for the formation of U141. 
Although the double cores in U141 might be explained with star forming \hii regions,
  existence of the \hii regions do not simultaneously explain the boxy shapes. 
In the case of a merger of dwarf galaxies, double cores and boxy shape of U141
  can be simultaneously explained. 
Although it is expected that low mass galaxies experience less merging events \citep{Lucia06},
  examples of dwarf-dwarf merger have been multiplying significantly in recent studies
  (\citealt{Del12}; \citealt{Amo14}; \citealt{Paudel15}).

During the merger of galaxies, gas supply toward the center becomes extremely efficient,
  producing a central star burst \citep{Bek08, Pee09}.
When the merger-induced central star burst is triggered, the galaxy will look like
  a blue centered galaxy. 
In addition, it is expected that the central gas metallicity will be lower than that
  expected from the luminosity-metallicity relation,
  due to the pure gas infall \citep{Pee09}. 
Star formation rate also will be larger than that expected from the SFR-mass relation \citep{Pee09}. 
However, U141 has higher gas metallicities (12 $+$ log(O/H) $=$ 8.4 and 8.6) for the two cores and
  lower star formation rate (log SFR [M$_\odot$ yr$^{-1}$] $\sim -$2.85)
  than the expected metallicity (12 $+$ log(O/H) $\sim$ 8.0) and 
  the star formation rate estimated from the SFR-mass relation (log SFR $\sim -$1.6), respectively. 
U141 is similar to the galaxies with high gas metallicity and low SFR found by \citet{Pee08}. 
They predicted that those are galaxies with depleted \hi compared to the abundance of oxygen
   in the final stages of their star forming activities.
Within this context, we suspect that U141 might be formed from a merger
  between \hi depleted galaxies
  having overall low star formation rate in the end of their star formation activity. 
During the merger of these \hi depleted progenitors,
  the star formation efficiency of U141 might has been low and gas metallicity been high.

If U141 is formed by a merger, we can roughly estimate the time scale of merging
  between two galaxies for which we assume a mass ratio of 2:1
  between primary and secondary based on the mass ratio of the two cores. 
We use the equation (4) from \cite{Jia08} for the dynamical friction time scale,
  where the secondary starts to merge from the virial radius of the primary
  with an elliptical orbit ($\epsilon =$ 0.5) and a relative radial velocity of 158 km s$^{-1}$
  which is the velocity dispersion of the Ursa Major cluster. 
The virial radius is converted from the effective radius using the relation
  between virial radius and effective radius \citep{Kra13}. 
We obtain a merging timescale of $\sim$1 Gyr for the two progenitors
  to arrive at the separation of $\sim 312$ pc and
  the relative radial velocity of $\sim 3$ km s$^{-1}$. 
In the end, the cores will be merged within $\sim$250 Myr.  

Nevertheless, the shallow SDSS images may prevent us from excluding other possible scenarios,
  particularly tidal interaction with other nearby galaxies. 
We cannot entirely rule out the possibility that U141 is in the stage of
  ongoing interaction with either UGC 7176 or NGC 4157. 
Given the very interesting nature of U141, we further suggest that
  a deep imaging and kinematic study would be extremely valuable.

\section*{Acknowledgments}
We would like to thank the anonymous referee for the very helpful comments and clarifications
  that helped to improve the manuscript.
Y. L. was supported by the Korea Astronomy and Space Science Institute (KASI)
  under the R\&D program supervised by the Ministry of Science, ICT and Future Planning.

This study is based on the archival images and spectra from the Sloan Digital Sky Survey (the full acknowledgment can be found at http://www.sdss.org/collaboration/credits.html) and  NASA Galaxy Evolution Explorer (GALEX). Funding for the SDSS has been provided by the Alfred P. Sloan Foundation, the Participating Institutions, the National Science Foundation, the U.S. Department of Energy, the National Aeronautics and Space Administration, the Japanese Monbukagakusho, the Max Planck Society, and the Higher Education Funding Council for England. The SDSS Web Site is http://www.sdss.org/. GALEX is operated for NASA by the California Institute of Technology under NASA contract NAS5-98034. We also acknowledge the use of NASA's Astrophysics Data System Bibliographic Services and the NASA/IPAC Extragalactic Database (NED).




\end{document}